\documentstyle[twocolumn,aps]{revtex}
\newcommand{\be}{\begin{equation}}
\newcommand{\ee}{\end{equation}}
\newcommand{\bea}{\begin{eqnarray}}
\newcommand{\eea}{\end{eqnarray}}
\newcommand{\bes}{\begin{eqnarray}}
\newcommand{\ees}{\end{eqnarray}}
\newcommand{\no}{\noindent}

\begin{document}
\draft
\title{ A unified representation of the q-oscillator and the q-plane}
\author{T.Rador\thanks{e-mail: rador@buphy.bu.edu}}
\address{Boston University, Department of Physics\\
Boston MA 02215, USA}
\maketitle
\begin{abstract}
Using deformations inspired by relativistic considerations and phase space
symmetry, we deform the position and momentum operators in one dimension.
The resulting algebra is shown to yield the q-oscillator algebra in one 
limiting case and the q-plane commutation relation in 
another.
\end{abstract}

In 1992 Arik and Mungan \cite{art42} 
anticipated a connection between the q-oscillator and relativity, as was
previously done by R.M.Mir-Kasimov \cite{art44} 
in a different context. Quoting from
Arik and Mungan, the momentum position commutator of a n-dimensional 
q-oscillator is given by 

\be{\label{eq:IIIarik1}}
\left[ P_{j},Q_{j} \right] = -i\left(1+\frac{q^{2}-1}{q^{2}+1}\sum_{k=1}^{j}\left(P_{k}^{2}+Q_{k}^{2}\right) \right) \; .
\ee

Here $P$'s and $Q$'s are dimensionless quantities. To express them
in terms of the physical position and momentum operators we perform the
usual replacement

\be{\label{eq:IIIarik2}}
Q_{k}\to\sqrt{m\omega/\hbar}Q_{k}\; , P_{k}\to\sqrt{m\omega\hbar}P_{k}\; .
\ee

If we also choose the dimensionless variable $q$ as

\be{\label{eq:IIIarik3}}
q=1+\frac{\hbar\omega}{mc^{2}}\cdots\; ,
\ee

\no where the omitted terms denote higher powers of $\hbar\omega/mc^{2}$, it
is evident that the contraction can be accomplished in different ways. Letting
$q\to 1$ as $c\to\infty$ in (\ref{eq:IIIarik1}), we obtain the Heisenberg
algebra suggesting that the algebra (\ref{eq:IIIarik1}) is relativistic. 
Another contraction is the limit $q\to 1$ as $\omega\to 0$, which corresponds
to turning off the oscillator interaction. In view of the former we expect
that this will yield a relativistic free particle algebra. In the latter
limit the commutator (\ref{eq:IIIarik1}) becomes

\be{\label{eq:IIIarik4}}
\left[ P_{j},Q_{j} \right] = -i\hbar\left(1+\sum_{k=1}^{j}\left(\frac{P}{mc}\right)^{2} \right) \; .
\ee

In the remaining part of their paper Arik an Mungan have proved that
this relation is indeed a free relativistic algebra. 

The one dimensional case of (\ref{eq:IIIarik4}) is 

\be{\label{eq:IIIkirkbes}}
\left[ P,X \right] = -i\hbar\left(1+\frac{P^{2}}{m^{2}c^{2}}\right) \; .
\ee

This commutator can be viewed as a definition for $P$ if we are
given $X=x$. The conclusion is that deforming the momentum
operator and leaving the position operator undeformed yields 
an algebra which describes a relativistic free motion.

We now deform the position operator in the same way as we 
have deformed the momentum operator 

\bes{\label{eq:IIIkirkalti}}
X &=& {\hbar \over \tau} Sinh ( {\tau \over \hbar} x) \; , \\ 
P &=& {\hbar \over \delta} Sinh(-i \delta D) \; .
\ees 

\no where the parameter $\tau$ is a free parameter having 
the dimensions of momentum. We point out that the deformation of 
the position operator in  is based on the same footing as the 
deformation of the momentum operator in , that is both can
be considered as a difference operator in the imaginary direction.
Thus the symmetry between $x$-space and $k$-space is maintained. 
Now we introduce two dimensionless constants $\mu$ and $\nu$ by

\bea
\delta &=& \mu {\hbar / mc}  \; , \nonumber\\
\tau &=& \nu mc \; . 
\eea

It can be shown that 
the commutation relation of the 
deformed position and the deformed momentum  operator in 
(\ref{eq:IIIkirkalti}) is 

\begin{eqnarray}\label{eq:IIIkirkyedi}
[P,X]=&&-i\hbar{Sin(\mu\nu)\over{\mu\nu(1+Cos\mu\nu)}}\times \nonumber \\
 && \left\{ \sqrt{1+\mu^2 ({P \over mc})^2} ,\sqrt{1+\nu^2 ({mc X \over \hbar})^2} \right\} \; ,
\end{eqnarray}

\noindent where \{a,b\} is the anti-commutator. We will now consider two
limiting cases of the algebra defined by (\ref{eq:IIIkirkyedi}). 
Although we have used 
the definitions (\ref{eq:IIIkirkalti}) to calculate the commutation relation 
(\ref{eq:IIIkirkyedi}), we can disregard them when considering the limiting 
cases we have claimed.  

An  expansion to the first non trivial order of (\ref{eq:IIIkirkyedi}) around $\mu^2 =0$, $\nu^2 =0$ yields,

\begin{equation}\label{eq:IIIkirksekiz}
[P,X]=-i\hbar \left(1+{1\over 2} ({\mu \over mc})^2 P^2 +{1\over 2}
({\nu mc \over \hbar})^2 X^2 \right).
\end{equation}

The commutator (\ref{eq:IIIkirksekiz}), to lowest order in $\mu$ and $\nu$, is the same as the q-oscillator commutation 

\begin{equation}\label{eq:IIIkirkdokuz}
[P,Q]=-i\left(1+{q^2-1 \over q^2+1}(P^2+Q^2)\right).
\end{equation}

Here $P$ and $Q$ are the dimensionless momentum and
position operators. We can embed the oscillator interaction by letting
$Q\to ({m\omega/\hbar})^{1/2}Q$ and $P\to ({mw\hbar})^{-1/2}P$. Comparing
the two commutators (\ref{eq:IIIkirksekiz}) and (\ref{eq:IIIkirkdokuz}) after 
this replacement we find the 
correspondence relations

\bea{\label{eq:IIIelli}}
{\nu \over \mu} & = &  {\hbar \omega \over m c^2} \; , \nonumber \\
q & = & 1+{\mu \nu \over 2} + \cdots \; .
\eea

The $\nu/\mu$ in (\ref{eq:IIIelli}) is the usual dimensionless
quantity encountered in the parameterization of q. As we have noted Arik
and Mungan showed that
by identifying $q=1+{\hbar\omega/ mc^2}$ and
considering the limit $\omega\to0$ one gets a relativistic free particle 
algebra. This limit corresponds to turning off the oscillator interaction. 
Furthermore we see that $\omega$ is proportional to $\nu$, 
that is the $\omega\to 0$ limit can be achieved by letting $\nu\to 0$. 
Therefore we can argue that the deformation
of the momentum operator alone yields a relativistic free particle algebra,
in addition deforming the position operator yields an interacting system with
interaction parameter $\omega$ being proportional to the deformation parameter
$\nu$.

Another interesting limit of (\ref{eq:IIIkirkyedi}) 
can be calculated by 
taking $\nu=\beta \sqrt{\alpha + 2\pi n}$, $\mu=\beta^{-1} \sqrt{\alpha +
2\pi n}$ and letting $n\to\infty$. Here $\alpha$ and $\beta$ are arbitrary. 
This results in

\no\hrulefill
\setlength{\unitlength}{0.012500in}%
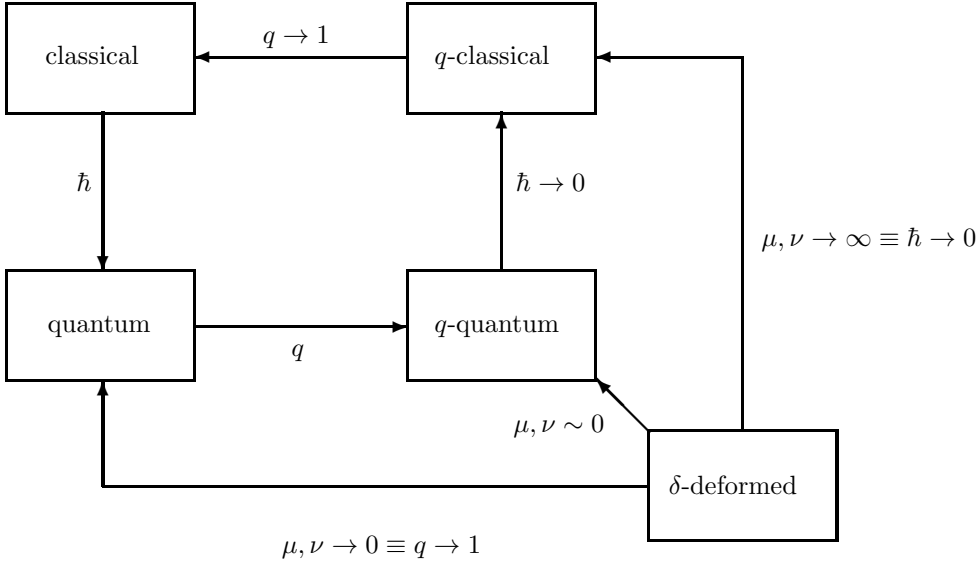
\begin{figure}{\centering{\begin{picture}(347,224)(15,601)
\thicklines
\put( 15,780){\framebox(78,45){}}
\put( 15,668){\framebox(78,45){}}
\put(183,668){\framebox(78,45){}}
\put(183,780){\framebox(78,45){}}
\put(284,601){\framebox(78,45){}}
\put( 55,780){\vector( 0,-1){ 67}}
\put( 94,690){\vector( 1, 0){ 89}}
\put(222,713){\vector( 0, 1){ 67}}
\put(183,803){\vector(-1, 0){ 90}}
\put(284,623){\line(-1, 0){229}}
\put( 55,623){\vector( 0, 1){ 45}}
\put(323,646){\line( 0, 1){157}}
\put(323,803){\vector(-1, 0){ 62}}
\put(284,646){\vector(-1, 1){ 22.500}}
\put(194,800){\makebox(0,0)[lb]{\smash{$q$-classical}}}
\put(292,620){\makebox(0,0)[lb]{\smash{$\delta$-deformed}}}
\put( 31,800){\makebox(0,0)[lb]{\smash{classical}}}
\put( 32,688){\makebox(0,0)[lb]{\smash{quantum}}}
\put(194,688){\makebox(0,0)[lb]{\smash{$q$-quantum}}}
\put(122,809){\makebox(0,0)[lb]{\smash{$q\to1$}}}
\put(134,678){\makebox(0,0)[lb]{\smash{$q$}}}
\put( 44,747){\makebox(0,0)[lb]{\smash{$\hbar$}}}
\put(228,747){\makebox(0,0)[lb]{\smash{$\hbar\to0$}}}
\put(130,595){\makebox(0,0)[lb]{\smash{$\mu,\nu\to0\equiv q\to1$}}}
\put(227,646){\makebox(0,0)[lb]{\smash{$\mu,\nu\sim0$}}}
\put(331,724){\makebox(0,0)[lb]{\smash{$\mu,\nu\to\infty\equiv\hbar\to0$}}}
\end{picture}}\vspace*{3ex}\caption{The diagram of $\hbar$, $q$ and $\delta$ deformations}
\label{fig:IIIyedi}}\end{figure}

\bea{\label{eq:IIIellibir}}
P X = q X P \; , \nonumber \\
q=(1+e^{-i \alpha})/(1+e^{+i \alpha}) \; .
\eea

The relations (\ref{eq:IIIellibir}) 
describes the generators that form the
classically deformed phase space ${R^2}_q$. Interestingly we 
obtain relation (\ref{eq:IIIellibir})  after "quantization" 
whereas in \cite{art43} it is postulated as a q-deformation of the 
classical $R^2$ without quantization. This is not a contradiction at all 
since in (\ref{eq:IIIellibir}) $\hbar$ has disappeared. 
This deformation limit can also be obtained by taking $\hbar\to0$ 
which turns off the quantization. Therefore we can argue that relation
(\ref{eq:IIIkirkyedi}) 
is an example of the "diagram" (see FIG.\ref{fig:IIIyedi}) 
of $\hbar$ and q deformations
presented in \cite{art43}, since we have obtained the q-deformed classical 
relation (\ref{eq:IIIellibir}) from a $\delta$-deformed quantum relation 
(\ref{eq:IIIkirkyedi}) by taking $\hbar\to0$. Both relations 
(\ref{eq:IIIkirksekiz}) and 
(\ref{eq:IIIellibir}) allow us to state that the difference operator and the 
deformed position operator  are closely related to 
q-deformations and to relativity.

Thus we have, at least in one dimension shown that a close 
connection exists between relativity, q-oscillators and difference 
operators. This work is based on a previous unpublished preprint by
the author and M.Arik {\cite{biz}}.


\begin{references}
\bibitem{art42}
Arik M. and Mungan M.
\newblock {\em Phys. Lett. B}, Vol.282, 101, 1992.

\bibitem{art44}
Mir-Kasimov~R. M.
\newblock {\em J.Phys. A: Math. Gen.}, Vol.24, 4283, 1991.

\bibitem{art43}
Aref'eva~I. Ya. and Volovich~I. V.
\newblock {\em Phys. Lett. B}, Vol.268, 179, 1991.

\bibitem{biz}
M.Arik and T.Rador, BUFB-94-01b, Bo\~{g}azi\c{c}i University Preprint. 

\end{references}
\end{document}